# Time-distance vision transformers in lung cancer diagnosis from longitudinal computed tomography

Thomas Z. Li.[a,b], Kaiwen Xu[c], Riqiang Gao[c], Yucheng Tang[d], Thomas A. Lasko[c,e], Fabien Maldonado[f], Kim Sandler[g], Bennett A. Landman [a,c,d,g]

[a]Biomedical Engineering, Vanderbilt University, Nashville, TN, USA 37235; [b]School of Medicine, Vanderbilt University, Nashville, TN, US 37235; [c]Computer Science, Vanderbilt University, Nashville, TN, USA 37235; [d]Electrical and Computer Engineering, Vanderbilt University, Nashville, TN, USA 37235; [e]Biomedical Informatics, Vanderbilt University, Nashville, TN, USA 37235; [f]Medicine, Vanderbilt University Medical Center, Nashville, TN, USA 37235; [g]Radiology & Radiological Sciences, Vanderbilt University Medical Center, Nashville, TN, USA 37235;

## ABSTRACT

Features learned from single radiologic images are unable to provide information about whether and how much a lesion may be changing over time. Time-dependent features computed from repeated images can capture those changes and help identify malignant lesions by their temporal behavior. However, longitudinal medical imaging presents the unique challenge of sparse, irregular time intervals in data acquisition. While self-attention has been shown to be a versatile and efficient learning mechanism for time series and natural images, its potential for interpreting temporal distance between sparse, irregularly sampled spatial features has not been explored. In this work, we propose two interpretations of a time-distance vision transformer (ViT) by using (1) vector embeddings of continuous time and (2) a temporal emphasis model to scale self-attention weights. The two algorithms are evaluated based on benign versus malignant lung cancer discrimination of synthetic pulmonary nodules and lung screening computed tomography studies from the National Lung Screening Trial (NLST). Experiments evaluating the time-distance ViTs on synthetic nodules show a fundamental improvement in classifying irregularly sampled longitudinal images when compared to standard ViTs. In cross-validation on screening chest CTs from the NLST, our methods (0.785 and 0.786 AUC respectively) significantly outperform a cross-sectional approach (0.734 AUC) and match the discriminative performance of the leading longitudinal medical imaging algorithm (0.779 AUC) on benign versus malignant classification. This work represents the first self-attention-based framework for classifying longitudinal medical images. Our code is available at https://github.com/tom1193/time-distance-transformer.

**Keywords:** Lung Cancer, Pulmonary Nodules, Longitudinal Vision Transformer, Time-Distance Vision Transformer, Temporal Emphasis Model, Longitudinal CT

## 1. INTRODUCTION

With lung cancer being the leading cause of cancer death and a low prevalence of malignancy in pulmonary nodules (1.1% to 12%)[1], the management of screen-detected pulmonary nodules has become a substantial public health problem. Although lung cancer screening with annual computed tomography is now the standard of care for high-risk patients, the absence of highly accurate and noninvasive diagnostics leads to increased costs, anxiety, morbidity from unnecessary diagnostic procedures of benign lesions, and death from missed malignancy[2,3]. As such, the use of predictive models to distinguish malignant from benign pulmonary nodules is an area of intense investigation[4].

Deep learning approaches have been moderately successful in estimating the probability of malignancy from medical images. Liao et al.[5] utilized a two-step strategy of first identifying pulmonary nodules with a region proposal network and second estimating their malignant potential with a leaky noise-or network. However, this approach in modeling cross-sectional CT scans is different than a clinican's approach to pulmonary nodules as consideration of how nodule features change over time, when possible, is highly predictive of malignancy[3,6]. In lung cancer specifically, studies have consistently found nodule growth rate to be most closely associated with malignancy out of all studied features[7].

Several studies have attempted to model dynamic nodule changes over repeated imaging. Ardila et al.[8] integrated consecutive chest CTs in a multi-channel input to diagnose lung cancer. Wang et al.[9] presented an adaptive convolutional neural network (CNN) to segment lung tumors from consecutive MRIs to analyze geometric tumor changes. Gao et al.[10] leveraged time intervals between scan acquisitions in a long short-term memory network with time-distanced gates (DLSTM) to outperform the leading cross-sectional approach on a lung screening cohort. The DLSTM is particularly practical in realistic clinical settings because screening chest CTs are often acquired irregularly. Patients commonly miss annual screenings and providers often choose to acquire follow-up imaging on a more frequent schedule if a nodule is concerning[11]. In this setting, the time interval between scans is important for a personalized approach to malignancy risk because a hallmark of malignant nodules is their rapid growth rate whereas benign nodules are relatively stable over a long period (Figure 1).

Recent advances have established self-attention[12] as a high performing and versatile learning mechanism across many domains. Transformer-based approaches are state-of-the-art for longitudinal time series[13] and have outperformed once-dominant convolutional approaches in natural video learning on large datasets[14,15]. In the medical imaging domain, transformer models have been performant in 3D multi-organ segmentation, disease classification, and medical image reconstruction to name a few[16,17]. Another promising line of research is joint vision-language transformers, which have greatly advanced the processing of multimodal data[18,19]. Longitudinal and multimodal models have clear applications in pulmonary nodule evaluation because integrating longitudinal imaging and non-imaging clinical features are important for a personalized approach. Motivated by the success of transformers in longitudinal natural data and medical imaging applications, we ask if self-attention is an effective learning mechanism for sparse, irregularity sampled medical images.

This paper introduces two novel extensions, time-encodings and time-aware, on the standard vision transformer[20] (ViT) and evaluates their potential for modeling sparse, irregularly repeating medical images. The time-encoding ViT (TeViT) incorporates time-distance into standard positional encoding[12] which is added to the imaging features. This approach reflects a common strategy of encoding continuous time that has been successful in modeling time-series[13,21], but has not been evaluated on longitudinal vision tasks. Our time-aware ViT (TaViT) is an extension on the distance-aware transformer[22], which uses continuous distances between tokens to adjusts weights from the self-attention query and key product in each head. We innovate on this strategy by leveraging a temporal emphasis model to explicitly describe time dynamics and scale self-attention weights. This approach incorporates domain-specific inductive bias by explicitly modeling the importance of recently versus distantly acquired images. We use a dataset of synthetic nodules to specifically test our approaches' capacity for modeling time-distance between longitudinal scans. Finally, we compare our approaches to the leading longitudinal and cross-sectional imaging approach in benign versus malignant classification of a lung screening chest CT cohort.

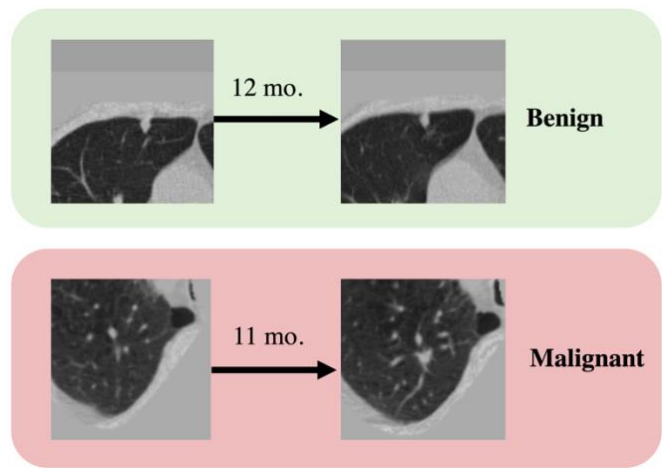

Figure 1. Examples of benign and malignant pulmonary nodules that would be challenging to classify without repeated imaging and consideration of the time interval between scans. The upper row shows axial regions of interest (ROIs) from two consecutive lung screening CTs depicting a benign nodule that exhibits minimal change over one year. The ROIs in the lower row are from a patient who developed lung cancer, where a malignant nodule grew aggressively over 11 months. Crops were made after sampling from the same resolution

## 2. METHODS

In sections 2.1 and 2.2 we introduce two versions of a ViT for modeling irregularly repeating images. The TeViT encodes a vector representation of time-distance between tokens, which is then added to the input tokens. The TaViT, scales self-attention weights with a temporal emphasis model.

### 2.1 Time Encoding ViT

Given a sequence of $T$ images, $\{x_1, ..., x_T\}$, and a sequence of corresponding image acquisition times, $t = [t_1, ..., t_T]$, we extract $N$ total feature tokens in $\mathbb{R}^D$ using separate strategies for 2D images and 3D whole chest CTs. For 2D images, we extract uniform patches from each image and project them into $D$-dimensional latent space with a learnable linear transformation, an embedding strategy that is standard in vision transformers. With this strategy, $N$ is $T$ times the number of patches extracted from each image. For 3D chest CTs, we leverage a pretrained nodule detection model to extract features from suspicious nodule regions. These features are linearly projected into a $D$-dimensional latent space with a learnable linear transformation. In this way, $N$ is $T$ times the number of proposed regions from each image. For each subject we construct a relative time distance vector $r \in \mathbb{R}^T$ where $r_t = |t_T - t_t|$. In the Time Encoding ViT, we generalize the encoding scheme first proposed in Vaswani et al.[12] to continuous time distances. A time encoding, $TE$, is constructed with sinusoids at different frequencies:

$$TE(r_t)[2i] = \sin\left(\frac{r_t}{10000^{\frac{2i}{D}}}\right)$$

$$TE(r_t)[2i+1] = \cos\left(\frac{r_t}{10000^{\frac{2i}{D}}}\right)$$

where $i = 0, ... D/2 - 1$ and $TE(r_t) \in \mathbb{R}^D$. This encoding scheme incorporates linear time distance information, since at any sampling time $t$ with time encoding $TE_t$ there is a linear transformation to the time encoding $TE_{t+k}$ at sampling time $t+k$. We construct $T$ unique time encodings and add these to tokens of the corresponding temporal index. The sequence of time encodings + feature tokens is input into a standard multi-headed self-attention encoder (Figure 2A).

### 2.2 Time Aware ViT

The TaViT aims to learn a temporal emphasis model (TEM) to scale self-attention at each head (Figure 2B). The purpose of the TEM is to represent the decrease in information over irregularly sampled time. This is a desirable quality in many medical imaging applications since data are acquired irregularly as cross-sectional manifestations of a continuously progressing phenotype. Intuitively, data sampled most recently to the time of diagnosis is closer to the underlying phenotype than is data sampled early on. The input consists of a $N \times D$ matrix $H = [\vec{h}_1; ...; \vec{h}_N]$, where $h$ is a token in $\mathbb{R}^D$ constructed with the same feature extraction schemes as described in section 2.1, and the acquisition times of each image, $t$. We construct a matrix of relative time distances $R \in \mathbb{R}^{N \times N}$ using the formula $R_{i,j} = |t_T - t_i| \; \forall \; j = 1, ..., N$. Here, $t_i$ is the acquisition time of the image from which token $h_i$ was derived from. Within each attention head, we learn a parameterized TEM that maps relative time distances in $R$ to non-negative scaled values, $\hat{R}_{i,j}$. There are many choices of TEMs[10] and we choose a flipped sigmoid function of the form:

$$\hat{R}_{i,j} = f(R_{i,j}) = \frac{1}{1 + \exp(aR_{i,j} - c)}$$

where $a$ and $c$ are non-negative learnable parameters that govern the slow of decline and time distance offset. This TEM has some desirable qualities that make it suitable for describing time dynamics. First, the maximum value of the function occurs at $R_{i,j} = 0$ for all non-negative $R_{i,j}$, which applies the largest scaling to tokens of the most recent temporal index. Second, when $R_{i,j} = 0$ the function's value does not fall below 0.5 for all non-negative learnable parameters $a$ and $c$. This ensures that self-attention between tokens of the latest temporal index and all other tokens are not reduced below a factor of 0.5. Third, the function monotonically declines sharply at a slope parameterized by $a$ after a time distance offset parameterized by $c$. In this way, self-attention between tokens of all temporal indices, except the latest, and all other tokens can be scaled down. Fourth, the function is non-negative for all $R_{i,j} \in \mathbb{R}$, ensuring that scaling with the TEM is unsigned. We learn separate parameters of the same TEM form at each attention head.

The following discussion describes the adjustments that the TaViT makes to its raw self-attention weights. Given $H \in \mathbb{R}^{N \times D}$ as the input to attention head $p$, three linear transformations are computed with head-specific weight matrices:

$$Q_p = H_p W_p^Q \qquad K_p = H_p W_p^K \qquad V_p = H_p W_p^V$$

where $Q_p$, $K_p$, $V_p$ are the query, key, and value matrices respectively of head $p$ from input $H$. In the standard Transformer, scaled dot product attention is formulated as

$$Attention_p = softmax(Q_p K_p^\top / \sqrt{d}) V_p$$

where $d$ is the dimension of the query and key matrices. In the Time Aware ViT, we propose to use TEM computed values in $\hat{R}$ to adjust the attention weights before softmax scaling. Generally our time-aware strategy intends to decrease the importance of weights computed between tokens of distant temporal indices towards the prediction. However, if the product of $Q_p K_p^\top$ is signed, downscaling may have the opposite effect of what is intended. To ensure that attention adjustment is done in an unsigned direction, we follow the strategy of Wu et al. and gate raw self-attention weights in the query-key product with a ReLU activation function. In summary, the final output of attention head $p$ is computed as:

$$Time\ Aware\ Attention_p = softmax\left(\frac{ReLU(Q_p K_p^\top) * \hat{R}}{\sqrt{d}}\right) V_p$$

As in the standard transformer, the outputs across all attention heads are concatenated and projected into the token dimension $D$ to reach the output of the multi-headed attention layer.

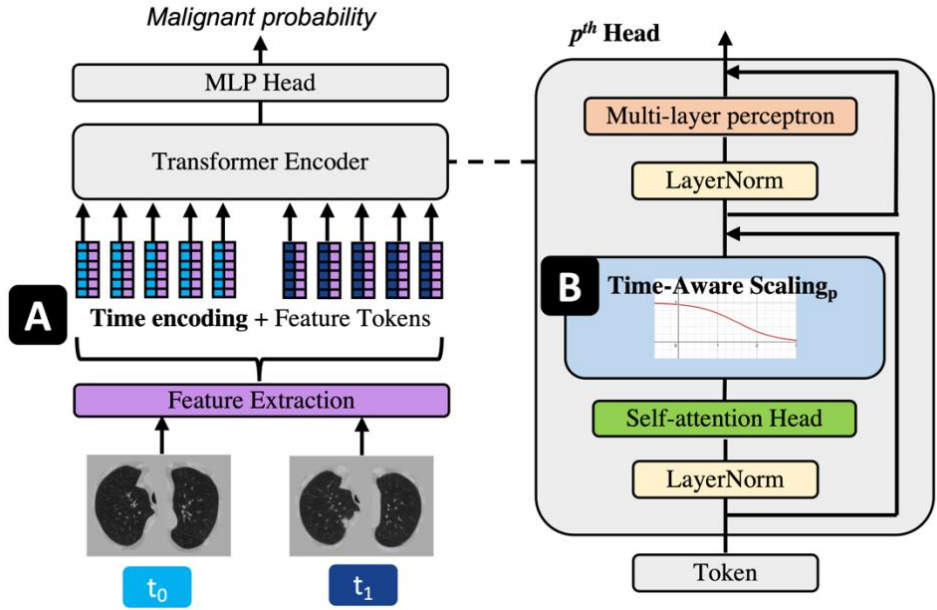

Figure 2. Feature extraction of 2D data was done by taking uniform patches and linearly projecting them in to a common embedding space. With 3D data, feature extraction is done with a region proposal network that proposes five local regions and embeds them as five feature tokens for each image volume. Time distance between repeated images are integrated into the model using two separate strategies. 2A: In the time embedding strategy, we add fixed vector representations of time distance, or time embeddings, to input tokens. 2B: In the time-aware strategy, we scale self-attention weights by a temporal emphasis model (TEM) in each attention head.

### 2.3 Baseline Models

We implement the Distanced Long Short Term Memory Network (DLSTM) as the leading longitudinal medical imaging algorithm. This model uses a TEM to encode time-distance into LSTM gates that govern information flow between hidden states. We compare our methods' results with the DLSTM on both synthetic and lung screening datasets. A positional ViT was also developed using positional encoding instead of the time distance strategies. The positional ViT was otherwise developed with the same training strategy and hyperparameters as the TeViT and TaVIT. We implement the lung nodule detection and cancer detection model from Liao et al. as a cross-sectional convolutional baseline, denoted as CS-CNN. Both methods have been validated on separate NLST cohorts for the task of discriminating malignant from benign subjects. We trained the CS-CNN and the DLSTM from random weights using our NLST cohort.

## 3. EXPERIMENTS

### 3.1 Synthetic Evaluation: Tumor-CIFAR

We used publicly available datasets Tumor-CIFAR v1 and v2 from Gao et al.[10] (https://github.com/MASILab/tumor-cifar) to run a proof of concept of our proposed methods. These contain 60,000 synthetic benign and malignant nodules superimposed on 32 x 32 CIFAR10[23] images across ten different classes. Each image in CIFAR10 is extended into 5 sequential images with growing nodules sampled from a simulation. In the simulation, the average growth rate of malignant nodules are three times that of benign ones, and both growth rates are sampled from gaussian distributions. In Tumor-CIFAR v1, the nodules are sampled 5 times at regular time intervals whereas in Tumor-CIFAR v2 they are sampled at irregular time intervals. We trained the TeViT, TaViT, positional ViT, and DLSTM with random cropping, horizontal flip, and intensity shift on benign versus malignant discrimination. Across both datasets, we compare the performance of these methods on a holdout test set.

Table 1. Cases and controls in the longitudinal NLST cohort with two consecutive lung screening chest CTs.

|  | # Subjects | # Scans |
| --- | --- | --- |
| Cases | 535 | 802 |
| Controls | 1397 | 2355 |

### 3.2 Screening Chest CT Evaluation: NLST

We curated a cohort of subjects with repeated lung cancer screening chest CTs from the NLST[24], a large randomized controlled trial that collected annual screening chest CTs from high-risk patients until they either detected lung cancer or the study period ended. We obtained two consecutive scans from each subject, taking the two most recent from those who had more than two scans. The inclusion criteria for cases were that the subject must have had a biopsy-confirmed diagnosis of lung cancer and that their latest scan was within 3 years of the diagnosis. The inclusion criteria for controls were subjects who were not diagnosed with lung cancer within the time frame of the clinical trial but had a positive screening result indicating the presence of a concerning lesion in the lung field. Since the controls greatly outnumbered cases, we randomly sampled from controls to achieve a 4:6 case control ratio. The final counts for our cohort are reported in Table 1.

Each chest CT were preprocessed with the following pipeline: resampling to 1 x 1 x 1 $mm^3$ isotropic resolution, segmentation of lung field, intensity windowing using a window of [-1200, 600] Hounsfield Units. We used the pretrained nodule detection model of Liao et al. (https://github.com/lfz/DSB2017) to infer local ROIs of pulmonary nodules from the preprocessed volumes. We extracted latent features of the five most confident ROIs from the pretrained model and these features were linearly projected into a 64-dimensional latent space with a learnable embedding, resulting in five input tokens at each time point. We perform a five-fold cross validation experiment with novel and baseline models and the reported performance metrics are the average of five models each trained and tested on a different fold.

### 3.3 Training Strategies

Tuning ViTs for various domain-specific tasks has been found to be nontrivial, and here we describe several training strategies that optimized model performance. First and foremost, pretraining with a masked autoencoding task[25] resulted in substantial increases in performance on both synthetic and empirical CT datasets. Importantly, masking is random across both subjects and time and the masking ratio increases with more time points, a scheme that has been shown to encourage learning of spatiotemporal dependencies[26]. We used a masking ratio of 0.75 in the Tumor-CIFAR dataset (5 sequential images) and 0.5 in the NLST dataset (2 sequential images). Second, using cosine warmup of learning rate[27] and AdamW optimization[28] led to a large speedup in convergence time of transformer models. Third, we added fixed 2D patch position embeddings[20] when evaluating on 2D images in Tumor-CIFAR. All transformer models in this paper were standardized to an embedding dimension $D$ of 64, a weight matrix dimension of 64, 8 attention heads, and an encoder depth of 8.

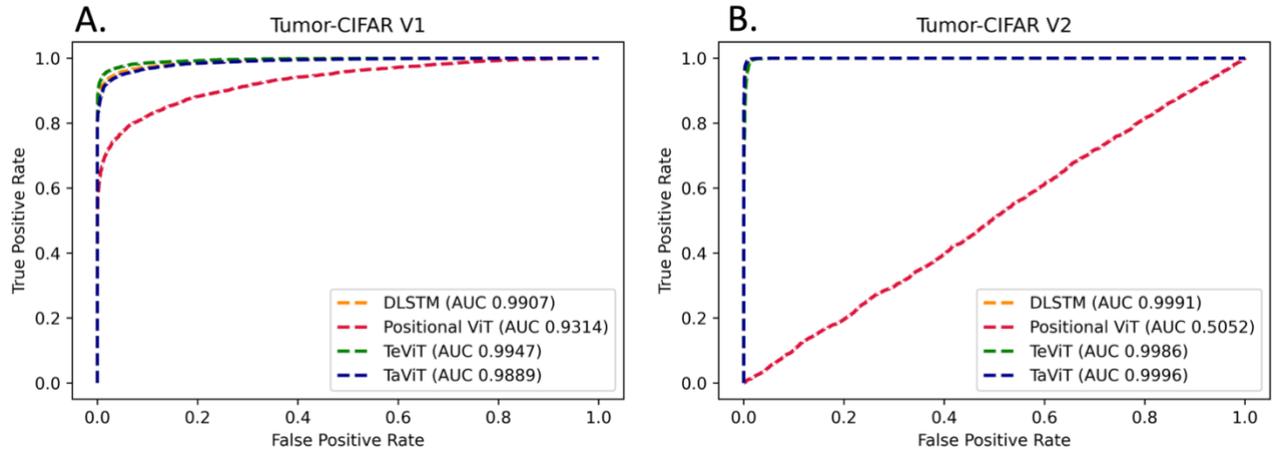

Figure 3. Receiver operating characteristic (ROC) curves of baselines and Longitudinal ViTs on Tumor-CIFAR. 3A: Both Longitudinal ViT approaches (TaViT and TeViT) slightly outperformed the baseline method (DLSTM) on regularly sampled synthetic nodules while the vanilla ViT performed slightly worse. 3B: The TaViT, TeViT, and DLSTM performed equally well on irregularly sampled synthetic nodules, while the positional ViT does not learn to distinguish classes at all.

## 4. RESULTS

### 4.1 Synthetic nodules: Tumor-CIFAR

As shown in Figure 3, the TaViT (AUC 0.9960) and TeViT (AUC 0.9948) slightly outperformed the positional ViT (AUC 0.9314) and DLSTM (AUC 0.9907) on version 1 of Tumor-CIFAR. Version 2 of Tumor-CIFAR samples nodules at irregular time intervals but the nodule size distributions of malignant and benign simulations are the same. On this dataset, The TaViT, TeViT, and DLSTM all achieved near perfect performance with AUCs of 0.9988, 0.9996, and 0.9991 respectively while the positional ViT was not discriminative at 0.5052 AUC.

### 4.2 Screening Chest CT: NLST

Table 2 presents the performance across five-fold cross validation on the longitudinal NLST lung screening cohort. Areas under the receiver operating characteristic curve (AUC) are reported as mean and standard deviation percentages from five cross-validated models tested on five different test sets. The TeViT and TaViT achieved a 0.05 improvement in average AUC over the CS-CNN. A two-tailed Wilcoxon signed-rank test found this performance difference to be statistically significant. The DLSTM was significantly more discriminative than the CS-CNN, recapitulating previous results on a different longitudinal NLST cohort. The TeViT and TaViT achieved a 0.02 improvement in average AUC over the positional ViT, but this improvement was not significant for $p<0.05$. All four longitudinal models achieved a significantly higher classification AUC than the CS-CNN.

Table 2. Percent average (standard deviation) of AUC across five-fold cross validation tests on the NLST cohort. P-value < 0.05 indicates that the TaViT significantly improves on the baseline methods using the Wilcoxon signed-rank test.

|  | AUC | p-value |
| --- | --- | --- |
| CS-CNN | 0.734 (0.249) | < 0.05 |
| DLSTM | 0.779 (0.214) | 0.75 |
| Positional ViT | 0.768 (0.348) | 0.06 |
| TeViT | 0.785 (0.225) | 0.81 |
| **TaViT** | **0.786 (0.332)** | **ref.** |

Revisiting the examples in Figure 1, Figure 4 compares the probability of cancer inferred from our time distance ViTs and the CS-CNN. The benign case depicts a large homogenous nodule without change over one year that the CS-CNN evaluates as having a malignancy risk of greater than 0.5 whereas our methods evaluate the risk to be less than 0.5. The malignant case depicts a smaller nodule that has grown over 11 months. In this example, the CS-CNN evaluates the cancer probability to be less than 0.5 whereas both time distance ViTs assign probabilities greater than 0.5.

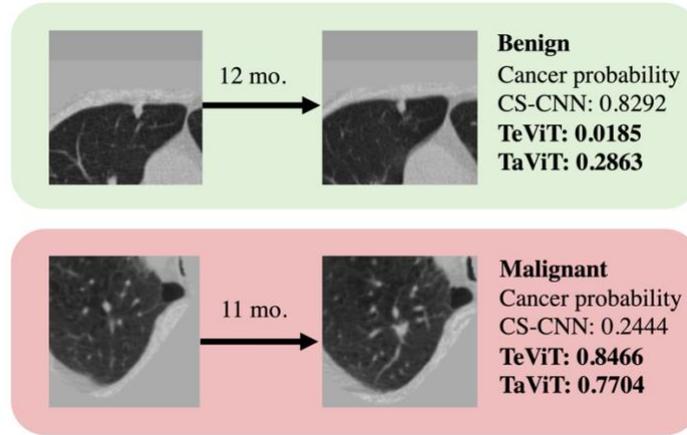

Figure 4 Cancer probabilities from the CS-CNN, TeViT, and TaViT for benign and malignant examples in Figure 1. In both examples the time distance ViTs are more accurate in estimating malignancy risk than the cross-sectional approach.

## 5. DISCUSSION

This work presents and validates the first techniques that integrate ViTs and time-distance to classify disease from irregularly repeating medical images. The TeViT adapts a time encoding technique that is well-validated in time series tasks. The TaViT proposes to directly scale raw self-attention weights with a temporal emphasis model. In this work, we use a standard ViT as the backbone architecture, but the proposed techniques can be integrated with more sophisticated transformer architectures as well.

The time encoding approach simply adds fixed vector representations of continuous time to input tokens. With the time-aware approach, we posit that self-attention between tokens of different temporal indices should be a function of the time distance between them. We design a temporal emphasis model that monotonically decreases as time-distance increases, thus allowing self-attention between tokens of distant temporal indices to be scaled down. Importantly, we learn a separate TEM in each attention head. As previously demonstrated in the distance-aware transformer, a subset heads may learn to heavily penalize on large time distances based on the features they attend to while others may learn to be time agnostic[22]. In both proposed methods, each token attends to all tokens from all temporal indices, which is in contrast to a LSTM approach where a cell state is sequentially updated through gated information flow[29].

Experiments on synthetic nodules show that the time distance ViTs have the capacity to classify irregularly sampled images, which is discussed further in section 4.1. We empirically validate our proposed methods on a screening chest CTs cohort and find that they are comparable with the DLSTM, with further discussion provided in section 4.2.

### 5.1 Synthetic experiment: Tumor-CIFAR

While both versions of Tumor-CIFAR were designed to evaluate learning of spatiotemporal features, there is much less variation in time intervals between repeated samples of the same subject in v1 than v2. Without the time-distance between repeated samples from the same subject, the positional ViT achieves minimal discrimination on v2. In contrast, the TeViT and TaViT achieve near perfect discrimination on v2, demonstrating the capacity to learn temporal dynamics that are associated with the classification task.

Our result on Tumor-CIFAR v1 again demonstrates the improvement of our methods compared to the positional ViT in modeling time-varying features. The fact that our methods achieve a similar performance with the DLSTM in v1 also suggests that they match the classification performance of convolution-based methods on a large dataset of 32x32 2D images, a result that has been reproduced in previous natural image classification studies[18,20,30]. Interestingly, pretraining our transformer-based methods with an asymmetric masked autoencoder was necessary to achieve DLSTM comparable performance on v1 but not on v2. While the importance of pretraining transformers is well known[19,20,31], our experiments suggest that pretraining specifically aids spatial learning more than temporal learning. The interpretation of these results are inherently limited by the programs used to generate the nodules, which assumes that nodule size, distribution, density, and growth rate come from a gaussian distribution. To provide empirical validation on clinically observed pulmonary nodules, we evaluate our methods on a cohort from the NLST.

## 5.2 Screening chest CT experiments: NLST

By achieved performance comparable with the leading DLSTM algorithm, our time distance ViTs are the first transformer-based algorithms to robustly incorporate time-distance information for performant longitudinal medical imaging classification. Our empirical validation was limited to a screening chest CT cohort, but our methods are easily adaptable to other longitudinal imaging tasks in the imaging and natural data domain. Both the time distance ViT and DLSTM significantly outperformed the CS-CNN, clearly indicating that the addition of prior scans improves benign and malignant discrimination of pulmonary nodules.

Interestingly, the time distance ViTs did not significantly improve on the positional ViT, reproducing findings of the DLSTM which also did not significantly outperform its positional equivalents on a slightly different longitudinal cohort of NLST. This result is counterintuitive as malignant tumors are known to grow much faster than benign nodules on average[3,7,32]. We offer three explanations for this finding. First, there are some lung cancer histologic subtypes such as indolent lung adenocarcinoma that have similar or even smaller growth rates as benign nodules[33–35]. This subtype may be overrepresented in our cohort, a possibility that warrants further analysis of histology reports. Second, there is certainly an overrepresentation of screen-detected cancers (cancers that occur during the screening period) in our longitudinal cohort because advanced stage cancers were intervened on after the subject's first scan and systematically excluded with our two-scan inclusion criteria[24]. Screen-detected cancers are observed early in their growth and thus different from cancers that have grown to an advanced stage before they are detected through screening. The discrimination between early, screen-detected cancers and benign lesions may be less reliant on time-dependent features. Third, we suspect that the lack of statistically significant improvement between time-distance and positional approaches may be explained by the small variation in scan acquisition in the NLST dataset. The adherence to an annual lung screening schedule in this clinical trial was much higher than in community lung screening programs (>90% versus 55% adherence respectively)[11,24]. Lower adherence would result in larger variations in an inter-scan time interval, which, in theory, is more difficult to approximate with a positional approach. Therefore, a limitation of this work is the need to validate our methods on irregularly repeating medical images from a real clinical setting as opposed to a clinical trial.

Another limitation of this study was the use of a region proposal network to extract features from detected nodules. The DLSTM used this pipeline as well. While the focus of this work was to develop techniques to specifically address variable time-distances in repeated medical images, the proposed techniques in this paper are limited in their capacity to learn from complete medical image volumes. Our future work will look to integrate time-distance techniques with recent advances in 3D ViTs to develop an end-to-end optimized network that combines nodule detection and cancer diagnosis from longitudinal imaging.

## 6. ACKNOWLEDGEMENTS

This research was funded by the NIH through T32 training grants of 5T32GM007347-41 and 5T32GM007347-42 as well as R01CA253923-0. This work was funded in part by NSF CAREER 1452485 and NSF 2040462. This research is also supported by ViSE through T32EB021937-07 and the Vanderbilt Institute for Clinical and Translational Research through UL1TR002243-06. We thank the National Cancer Institute for making available lung screening CTs from the NLST project. This work was facilitated by the open source contributions of PyTorch[36].